\documentclass[showpacs,aps,nofootinbib,nobibnotes,superscriptaddress,preprint]{revtex4-1}
\usepackage{amssymb}
\usepackage{graphicx,amsmath}
\usepackage{bm}
\usepackage{times}
\usepackage{blindtext}
\usepackage{xcolor}
\usepackage[strings]{underscore}

\def\expect#1{\left\langle \!#1 \!\right\rangle}
\def\braAket#1#2#3{\left\langle \!#1\! \left|\! #2\! \right|\! #3\! \right \rangle}

\begin{document}


\title{Performance analysis of an optically pumped magnetometer in Earth's magnetic field}

%
%
\date{\today }
\author{G. Oelsner}
\email{gregor.oelsner@leibniz-ipht.de}
\affiliation{Leibniz Institute of Photonic Technology, P.O. Box 100239, D-07702 Jena, Germany}
\author{V. Schultze}
\affiliation{Leibniz Institute of Photonic Technology, P.O. Box 100239, D-07702 Jena, Germany}
\author{R. IJsselsteijn}
\affiliation{Supracon AG, An der Lehmgrube 11, D-07751 Jena, Germany}
\author{R. Stolz}
\affiliation{Leibniz Institute of Photonic Technology, P.O. Box 100239, D-07702 Jena, Germany}



\begin{abstract}
We experimentally investigate the influence of the orientation of optically pumped magnetometers in Earth's magnetic field. We focus our analysis to an operational mode that promises femtotesla field resolutions at such field strengths. For this so-called light-shift dispersed $M_z$ (LSD-Mz) regime, we focus on the key parameters defining its performance. That are the reconstructed Larmor frequency, the transfer function between output signal and magnetic field amplitude as well as the shot noise limited field resolution. We demonstrate that due to the use of two well balanced laser beams for optical pumping with different helicities the heading error as well as the field sensitivity of a detector both are only weakly influenced by the heading in a large orientation angle range.
\end{abstract}
\maketitle

%


\section{Introduction}

Optically pumped magnetometers (OPMs) are, in principle, scalar-type quantum sensors for magnetic fields based on the Zeeman effect. That is the shift of energy levels due to the interaction of atoms with an external magnetic field \cite{CohenTan1977}. Usually alkali vapors in paraffin-coated glass cells are used as sensing element. Because the energy shift by the Zeeman interaction is based on the scalar product of the measured external magnetic field $\vec{B}_0$ and the magnetic moment of the atom, such a magnetometer measures the absolute value of the field. This fact makes them interesting for the realization of absolute field sensors \cite{Leger2009,Nabighian2010,Korth2016}.

On the other hand, the alkali vapor is usually polarized by optical pumping with a circular laser beam to create large signal amplitudes. With the pump beam an additional direction dependence is introduced leading besides dead zones also to unwanted effects usually summarized under the label "heading error"\cite{Budker2013}. These effects can be understood from the change in the atom-light coupling that is in dipole approximation given by the scalar product of the laser's electric field and the atoms electric dipole moment \cite{Oelsner2019}. Suppression of the heading error is intensively studied \cite{Jensen2009,Bao2018,Hu2018} and it has important consequences in the application of OPMs \cite{Colombo2016,Weis2017}.

The most sensitive magnetic field sensors are based on superconducting quantum interference devices \cite{SQUIDHB,Schmelz2017}. They have been optimized to allow sub-femtotesla field gradient resolution even in Earth's magnetic field \cite{Stolz2001} and today, besides others, are used for geomagnetic and archeological explorations \cite{Chwala2001,Foley2004,Stolz2015}. Still the requirements due to cryogenic liquids is demanding and for low temperature superconducting sensors the costs of liquid helium are high. In this context, the newly introduced operational modes which enable shot-noise-limited field resolutions of OPMs on the femtotesla scale in Earth's magnetic field strengths are promising alternatives. Namely, those are the light narrowing (LN) \cite{Scholtes2011,Guo2019,Fu2019} and the LSD-Mz \cite{Schultze2017} mode.

In this work, we analyze the performance of magnetometers based on the LSD-Mz mode and experimentally investigate their characteristics as a function of their orientation in an external magnetic field. The paper is organized as follows: We describe the experimental setup before we discuss a theoretical description of the measured values. Afterwards we present experimental data and relate it to our theoretical description. We demonstrate the influence of different rotational axis to the results. Finally, our model allows an estimate of the influence of the heading direction on the field resolution which we use to conclude on the usability of OPMs based on this mode in Earth's magnetic field.

\section{Methods}
\subsection{Experimental description}
\label{sec1}
\begin{figure}[h]
\includegraphics[]{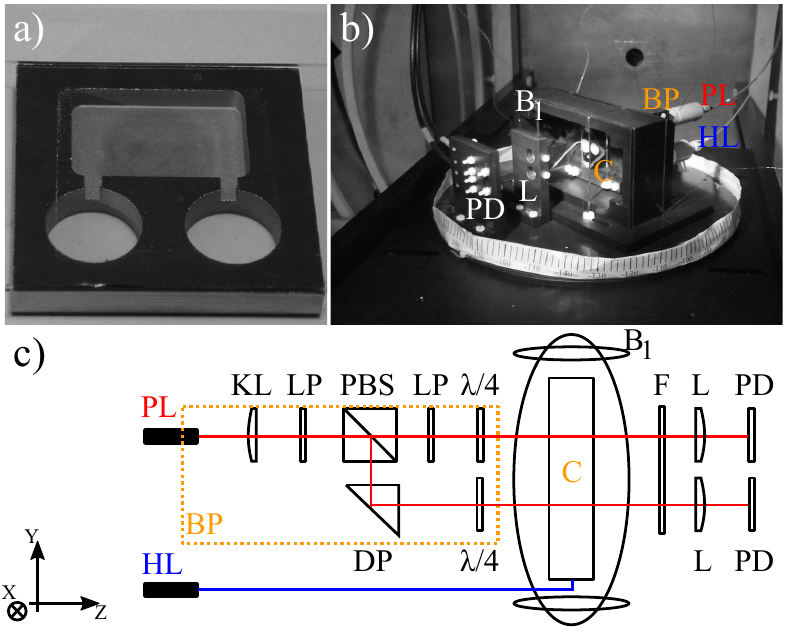}
\caption{Experimental setup. a) The vapor cell used for our experimental investigations features two circularly shaped active volumes and a rectangular reservoir. b) The experimental setup (see also Ref.~\cite{Oelsner2019}) is mounted on a rotational table inside of a Helmholtz coil system and a mu-metal shielding. The orientation of the setup concerning the created magnetic field is adjusted by a cable pull. c) The schematic drawing of the setup viewed from the side for rotating angles of $\alpha = 0$ includes the optics for generating two circularly polarized laser beams. The pump (PL) and heat laser (HL) are colored in red and blue, respectively. The whole optical setup can be rotated inside the magnetic field $\vec{B}_0$ that is fixed along the z-axis.} \label{Fig:setup}
\end{figure}

Because we are interested in the characteristics of a high resolution OPM as a function of its orientation relative to the Earth's magnetic field, our experimental setup is based on a vapor cell design suitable for the LSD-Mz regime. Namely, the used micro-fabricated buffer gas cell contains two active volumes connected to the same reservoir. It is created by ultrasonic milling the desired structure into a 4~mm thick silicon wafer. After filling the reservoir region with droplets of diluted cesium azide and a drying step, the cell is closed by anodic bonding with Borofloat glass plates. Finally, decomposing the azide to cesium and nitrogen as buffer gas with excimer laser irradiation finalizes the fabrication. A detailed description of the cell fabrication can be found in \cite{Woetzel2011}. A photograph of the used cell arrangement is shown in Fig.~\ref{Fig:setup}~a).

The cell is mounted at the center of a rotational table and accompanied by optic elements for the preparation and detection of the required circularly polarized laser beams, see Fig.~\ref{Fig:setup}~b). A schematic of the optical setup is shown in Fig.~\ref{Fig:setup}~c). The laser light is guided to the experiment by a polarization maintaining fiber. The beam preparation (BP) is done in a compact setup. Therein a lens collimates the beam to a diameter of about 4~mm and a linear polarizer (LP) is used to define the polarization direction. The beam is then split by the use of a polarizing beam splitter (PBS). A right angle deflecting prism (DP) and two quarter-lambda plates ($\lambda$/4) create two parallel beams with right and left handed circular polarization, respectively. An additional linear polarizer (LP) in the straight beam is used for balancing their intensities. The so prepared beams are passed through the active volume of the two cells, before they are slightly focused with lenses (L) onto two photodiodes (PD) for recording the transmitted intensities. Unwanted light on the diodes, as for example from the heating laser, is blocked with an absorbing bandpass filter F.

The pump laser (PL) with a wavelength of 895~nm is stabilized to the Doppler free absorption line for the $F=3 \rightarrow F^\prime$ = 4 D$_1$ transition of the Cs vapor of an additional paraffin coated glass cell. A 978~nm heat laser is guided to the setup by a fiber and an optical setup, containing collimating lens and deflecting prism (not show in Fig.~\ref{Fig:setup}). The sensor is heated to a temperature of roughly 100 degree Celsius.

The whole optical installation is mounted inside of a three layer mu-metal shielding (see Fig.~\ref{Fig:setup}b) and Ref.~\cite{Schultze2010}). Also, a three axis Helmholtz-coil system is included allowing the application of arbitrary magnetic fields. In all discussed experiments we apply a static magnetic field $\vec{B}_0$ of about $50$~$\mu$T along the z-axis that corresponds to the rotational axis of the cylinder-shaped shields. The normal of the rotational table and thus the rotational axis is labeled by y. Two additional Helmhotz coil configurations are mounted around the cell allowing for the application of magnetic rf-fields ($\vec{B}_1$) perpendicular to the laser light for the detection of the magnetic resonance. We respectively denote the amplitude and the frequency of the $B_1$ field by $\Omega$ and $\nu$. Note, the $\vec{B}_1$ field created by the y-coil is perpendicular to both, laser light direction $\vec{k}$ and magnetic field $\vec{B}_0$ for each rotation angle $\alpha$. But using the second rf-coil leads to an angle modification between $\vec{B}_0$ and $\vec{B}_1$ from perpendicular to parallel configuration during rotation. Nevertheless, we label the latter as "x-coil" according to its initial orientation. Our setup thus resembles the real situation of an OPM moved in Earth's magnetic field.

The measurements are carried out as follows: At first the heading angle is adjusted from outside of the shielding by a cable pull to an angle $\alpha$ between $\vec{B}_0$ and the propagation direction of the laser light $\vec{k}$. The latter is given by $\sin \alpha \vec{e}_x + \cos\alpha \vec{e}_z$. Here, the $\vec{e}_i$ denote the unit vectors in i-direction. Then, consecutively the $B_1$ field is swept through the magnetic resonance using the $x$ as well as $y$ coil. Thus, for each angle $\alpha$ recordings are made with a magnetic rf-field applied in $y$ and $\cos \alpha \vec{e}_x + \sin\alpha \vec{e}_z$ direction. Because at an angle $\alpha = 0$ both configurations are equivalent, we adjusted the current feed to the different $B_1$ coils to produce magnetic resonances with the same height and width at these angles to correct for the slightly different coil constants.

\begin{figure}[h]
\includegraphics[]{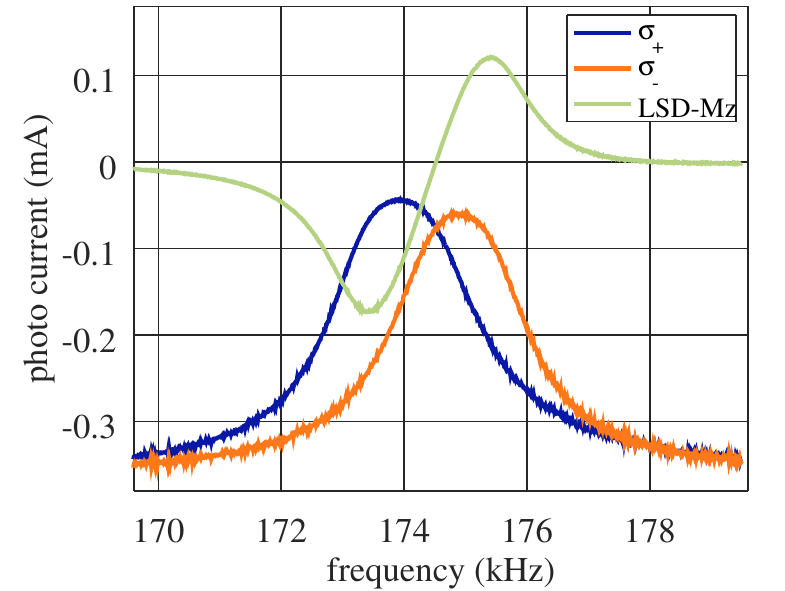}
\caption{Measured magnetic resonance for both helicities as well as difference signal. The latter yields a large transfer function exploited in the LSD-Mz mode. This example curves are measured at an angle $\alpha = 0$ with the y-coil.} \label{Fig:signals}
\end{figure}

In the starting configuration ($\alpha = 0$) the LSD-Mz mode requires two detuned circularly polarized laser beams both of which are oriented in parallel to the external magnetic field $B_0$ \cite{Schultze2017}. When using buffer gas cells, the population is pumped to dark states at $m_F = \pm 4$ of the $F=4$ ground states, respectively for $\sigma_+$ and $\sigma_-$ light. By the magnetic rf-field the transition between the Zeeman-split levels can be driven, which is observable by an increase in laser absorbtion. The substraction of the signals for both helicities results in a steep linear measurement curve around the actual Larmor frequency $\gamma B_0$. Here $\gamma = 3.5$~Hz/nT denotes the gyromagnetic ratio of cesium. This substraction also reduces the common noise present on the two signal, as for example intensity noise from the laser. For illustration, corresponding measurements of the magnetic resonances are presented in Fig.~\ref{Fig:signals} for an angle of $\alpha = 0$ using the y-coil.

By fitting this measurement curves of the photocurrent as a function of rf-frequency $I(\nu)$ with a Lorentzian function
\begin{equation} \label{Eq:Lorentzfit}
I = I(\nu) + I_{dc} = I_0 \frac{\Delta\nu^2}{4(\nu-\nu_0)^2+\Delta\nu^2} + I_{dc},
\end{equation}
we extract the DC-photocurrents $I_{dc}$, the amplitudes $I_0$ and widths (FWHM) $\Delta \nu$ of the magnetic resonances, as well as the resonant frequencies $\nu_0$. The latter is given by the light shifted Larmor frequency $\nu_0 = \nu_L \pm \nu_{LS}$, where $\nu_{LS}$ denotes the frequency shift due to the AC-Stark effect. Although the helicity of the two laser beams as well as their relative intensities were balanced to less than 1~\% prior to recording the experimental data, a remarkable deviation in both, the dc-current and the resonance amplitude, is observed. Therefore, in all measurements including the one presented in Fig.~\ref{Fig:signals} an additional electronic balancing through amplifying the $\sigma_+$ signal by an additional factor of $a = 1.61$ is introduced. The reason for the observed deviation, that is dependent on the applied magnetic field requires further investigations.

From the steepness of the difference signal $s = d(I_{\sigma_+} -I_{\sigma_-})/d\nu$ at the Larmor frequency we calculate the theoretical shot-noise limited sensitivity as
\begin{equation}\label{Eq:Bsn}
B_{sn} = \frac{\sqrt{2e \left(I_{dc}^{\sigma_+}(\nu_L)+ I_{dc}^{\sigma_-}(\nu_L)\right)}}{\gamma s} ,
\end{equation}
where the photocurrents $I_{dc}$ are taken at the Larmor frequency and $e$ is the elementary charge. The heading characteristics of the fitting parameters of Eq.~\ref{Eq:Lorentzfit} as well as of the sensitivity are in the following discussed.

\subsection{Theoretical description}
\label{sec2}

Two theoretical concepts are required for the description of our experimental results. On the one hand, the measured resonance frequencies are strongly modified by the light shift due to the intense off-resonant pumping. This effect has been extensively studied \cite{Oelsner2019,Jensen2009,Hu2018,Mathur1968} and we use the method presented in Ref.~\cite{Oelsner2019} for a description of our data.

On the other hand, the magnetic resonance itself can be described by Bloch equations. Here, we firstly analyze the Hamiltonian describing the magnetic resonance of a two-level system
\begin{equation}\label{Eq:Ham1}
H= \frac{h \nu_0}{2}\sigma_z + h\Omega\cos 2\pi\nu t \left[ \cos \alpha \sigma_x + \sin \alpha \sigma_z \right].
\end{equation}
for the x-coil. Please note that the same Hamiltonian is valid for the y-coil keeping $\cos\alpha = 1$ and $\sin\alpha = 0$. In Hamiltonian \eqref{Eq:Ham1}, $\Omega = \gamma B_1$ is introduced as the amplitude of the $B_1$ field in frequency units, $\sigma_x$, $\sigma_y$, as well as $\sigma_z$ denote Pauli matrices, and $h$ the Planck constant. In order to proceed, we aim to discuss the system in a rotating frame that removes the time dependent diagonal term. That can be achieved by a unitary transformation $U_1 = e^{iG(t) \sigma_z}$ where we choose $G(t) =\frac{\Omega}{\nu}\sin 2\pi\nu t \cos \alpha$. As constructed, $U_1$ commutes with $\sigma_z$ and the term $i\hbar \dot{U}_1 U_1^\dagger$ removes the diagonal coupling \cite{Shevchenko2014}. The transformation therefore results in
\begin{equation}\label{Eq:Hamprime}
H^\prime = \frac{h \nu_0}{2} \sigma_z + h \Omega \cos 2\pi\nu t \sin \alpha \left[ e^{ig\sin 2\pi\nu t} \sigma_+ + e^{-ig\sin 2\pi\nu t} \sigma_- \right].
\end{equation}
Here, we used $\sigma_\pm = 0.5 \left( \sigma_x \pm i \sigma_y \right)$ and the abbreviation $g = \frac{\Omega}{\nu}\cos \alpha$. Although we could continue with the Jacobi-Anger expansion and finally find the stationary terms in a frame rotating with $\nu$, we note that the ratio $\Omega/\nu$ is small in our experimental realization. Thus, we can summarize the lower line to $\sigma_x$ and solve the Bloch equations in a frame rotating with $\nu$ around the z-axis in rotating wave approximation. The modification of the stationary result for the z-component of the magnetization compared to the usual result is limited to a reduction of the effective rf-amplitude with a factor $\cos \alpha$ as
\begin{equation}\label{Eq:Blochres}
\expect{\sigma_z} = -\frac{\Gamma_r \Gamma_\varphi^\prime}{\Gamma_r \Gamma_\varphi^\prime + \Omega^2 \cos^2\alpha}.
\end{equation}
Above we introduced the rates $\Gamma_r$ and $\Gamma_\varphi$ as respected inverse relaxation and coherence times $T_1$ and  $T_2$, as well as $\Gamma_\varphi^\prime = (\Gamma_\varphi^2+\delta^2)/\Gamma_\varphi$. The detuning of the $B_1$ frequency from resonance is included by $\delta = \nu_0 - \nu$.

Still, this modification for the x-coil alone fails to accurately explain our measured data. What is missing in Eq.~\eqref{Eq:Blochres} is the modification of population transfer by optical pumping, or in other words the change in the dc-photocurrent that depends on the population difference. To include the optical transition and the decay from the atoms excited state into the two-level model, we modify the relaxation and excitation dissipative dynamics, similar as in Refs.~\cite{Dreau2011} and \cite{Shi2018}. Namely, we set for the respective excitation $\Gamma_e$ and decay rate $\Gamma_r$
\begin{equation} \label{Eq:moddiss}
\begin{split}
\Gamma_r &= \frac{\gamma_r}{2} + \frac{\Omega_p \left|\cos \alpha\right|}{2} \left(1-\cos \alpha \right), \\
\Gamma_e &= \frac{\gamma_r}{2} + \frac{\Omega_p \left|\cos \alpha\right|}{2} \left(1+\cos \alpha \right).
\end{split}
\end{equation}
Here we assumed a repopulation rate of the ground state levels $\gamma_r/2$ and an effective pumping rate between the two levels $\Omega_p \propto \Omega_L$. We note that the equations above are written for $\sigma_+$ polarization of the light. Still, they take the same form for $\sigma_-$ light. This other helicity only differs by a change of the signs inside the brackets. On the other hand, because the $m_F=-3$ level has higher energy compared to the $m_F = -4$ in two level approximation, the result for $\sigma_-$ would be equivalent to Eq.~\ref{Eq:moddiss}. The rates resulting from the optical pumping in Eq.~\ref{Eq:moddiss} include contributions of the scalar product between the atoms's dipole moment and the electric field ($\propto 1\pm \cos \alpha$) (compare to  Ref.~\cite{Oelsner2019}) as well as an additional factor of $\left|\cos \alpha\right|$. Also, we did not include the linear polarized component scaling with $\sin \alpha$ that, in principle, can also contribute to a population difference. This statement is especially true close to a perpendicular pumping orientation $\vec{B}_0 \perp \vec{k}$. The rates defined in \eqref{Eq:moddiss} enter to the observable photocurrent. It relates to the expectation value of $\sigma_z$ as
\begin{equation}\label{Eq:Blochres2}
I(\omega) = I_{0}\expect{\sigma_z} = I_{0}\frac{(\Gamma_e-\Gamma_r) \Gamma_\varphi^\prime}{(\Gamma_r+\Gamma_e) \Gamma_\varphi^\prime + c^2(\alpha)\Omega^2 },
\end{equation}
Here, the function $c(\alpha)$ is constantly one for the x-coil and $\cos \alpha$ for the y-coil. As constructed, our model accurately describes the dc-photocurrent as shown in Section~\ref{sec3}. Namely, if $\delta \rightarrow \pm \infty$ or equivalently $\Omega \rightarrow 0$, we find
\begin{equation}\label{Eq:zeropop}
I_{dc} = I_{0} \frac{\Gamma_e-\Gamma_r}{\Gamma_r+\Gamma_e }= I_{0}\frac{\cos^2 \alpha}{\left|\cos\alpha\right| + p_1},
\end{equation}
with the dimensionless fitting parameter $p_1 = \gamma_r/\Omega_p$. The above function changes with decreasing optical pumping from a $\cos$ to a $\cos^2$ angular dependence while at the same time the amplitude is decreased. This reduction of the dc-current summarizes the fact that for an effective pumping to the dark states the rate of equalization of population $\gamma_r$ should be significantly smaller than the optical pumping $\propto \Omega_p$.

To bring Eq.~\ref{Eq:Blochres2} into a similar form as Eq.~\ref{Eq:Lorentzfit} we separate the dc-current to find the Lorentzian-like resonance function as
\begin{equation}
I(\nu) = I_{dc}\left( 1- \frac{c^2(\alpha)\Omega^2}{\left(\Gamma_e+\Gamma_r \right)\Gamma_\varphi^\prime + c^2(\alpha)\Omega^2} \right).
\end{equation}
The resonance amplitude is accordingly the value of the second term in above equation at resonance, namely at $\nu = \nu_0$. It is
\begin{equation}\label{Eq:ampl}
i_{0} = I(\nu = \nu_0)/I_{dc} = \frac{c^2(\alpha)}{c^2(\alpha)+p_2\left( p_1+\cos\alpha\right)},
\end{equation}
where we introduced a second dimensionless fitting parameter $p_2 = \Gamma_\varphi \Omega_p / \Omega^2$.

Finally, the full width at half maximum (FWHM) of the resonance curve of Eq.~\eqref{Eq:Blochres2} is given by
\begin{equation}\label{Eq:width}
\Delta \nu = 2\Gamma_\varphi \sqrt{1+\frac{c^2(\alpha)}{p_2(p_1+\left|\cos \alpha \right|)}+ p_3}.
\end{equation}
This relation describes the quadratic addition of the natural linewidth $\Gamma_\varphi$ and the power broadening by the $B_1$ field
\begin{equation}\label{Eq:powerbroad1}
\Delta \nu_{B_1} = \frac{c^2 \Gamma_\varphi}{p_2\left(p_1+\left|\cos\alpha\right|\right))} = \frac{c^2\Omega^2}{\Gamma_\varphi \left( \gamma + \Omega_p \left|\cos\alpha\right| \right)}
\end{equation}
of the magnetic resonance line. To fit our experimental results, it is necessary to introduce an additional constant broadening term $p_3$ that we account to the broadening of the magnetic resonance line by the laser power as we will demonstrate below. We introduce this solely to the resonance width, since a saturation in the optical transition does not yield the same in the magnetic resonance and thus will not influence to the amplitude of the measured Lorentzian-shaped signal.

\section{Results}
\label{sec3}
\subsection{Resonant frequency - Light shift}
\begin{figure}[h]
\includegraphics[]{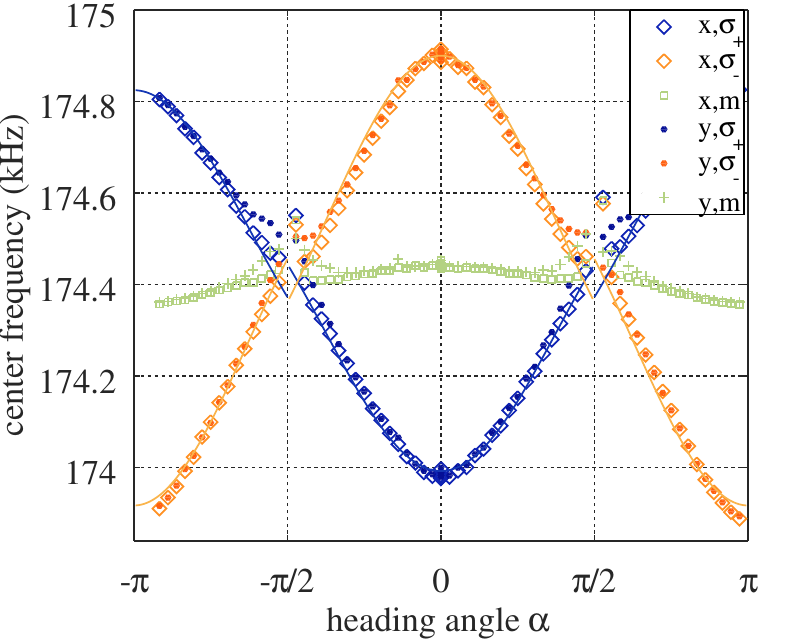}
\caption{Extracted center frequencies of the magnetic resonances for both circular polarizations of the laser as a function of the heading angle $\alpha$. Also the mean frequencies of the two helicities are given. The latter roughly corresponds to the Larmor frequency measured by the LSD-Mz mode. The diamonds and dots correspond to a $B_1$ field that is applied respectively in plane of rotation and perpendicular to it. Away from $\pm \pi/2$ the two coils produce the same result. Additionally, theoretical curves describing the expected light shift are added as solid orange and blue lines.} \label{Fig:freq}
\end{figure}

In a first step we analyze the angular dependence of the magnetic resonance center frequencies as plotted in Fig.~\ref{Fig:freq}. We added to the figure also the mean value of the measured center frequencies found for the two different circular polarizations.

In general, the magnetic resonances are strongly shifted by the ac-Stark shift due to the strong off-resonant pumping. The angle between the laser beam direction, represented by its $k$-vector, and the magnetic field $\vec{B}_0$ influences to the atom-light interaction by the transition dipole moment \cite{Oelsner2019}
\begin{equation}\label{Eq:dipole}
\vec{D}\cdot\vec{E} = \frac{E_0}{2\sqrt{2}} e^{i\left( \vec{k}\vec{r}-2\pi\nu_L t \right)} \left( \left[ \cos \alpha \pm 1 \right] D_+  + \left[ \cos \alpha \mp 1 \right] D_- +2 \sin \alpha D_z \right).
\end{equation}
Therein $E_0$ is the amplitude of the electric field, $\vec{k}$ the k-vector of the laser, $\vec{r}$ the atoms position, $\nu_L$ the frequency of the laser beam, and the $D_i$ are components of the dipole operator. In words, above equation states that with modifying the angle $\alpha$ the weight of pumping to different excited states is strongly influenced. Namely, $D_+$, $D_-$, and $D_z$ components couple to excited states with increased, decreased, as well as unchanged magnetic quantum number $m_F^\prime = m_F + \left\{+1,-1,0\right\}$, respectively.

The strongest characteristic results from the vector light shift that can be interpreted as virtual magnetic field added in direction of the lights angular momentum \cite{Mathur1968}. It is strongest close to angles of zero and $\pm\pi$ and reduces to zero close to perpendicular orientation. Also as expected, it changes its sign for a change in the laser's helicity. A more detailed discussion of the light shift can be found in Ref.~\cite{Oelsner2019}. We use the findings therein to calculate the expected light-shifted transition frequencies between Zeeman levels with highest and lowest magnetic quantum numbers to their neighboring states. These correspond to the respective blue and orange solid lines plotted in Fig.~\ref{Fig:freq} and can be identified with the transitions probed when pumping to the dark states $m_F = \pm 4$, where the plus and minus sign are to be used for $\sigma_+$ and $\sigma_-$ light, respectively. For the curves we used a magnetic field strength of $B_0 = 49.664$~$\mu$T, detunings of $\delta_{F^\prime = 4} = -8$~GHz and $\delta_{F^\prime = 3} = -9.2$~GHz from the respective optical transitions $F=4 \rightarrow F^\prime = 4$ and $F=4 \rightarrow F^\prime = 3$, a linewidth of the optical transitions $\Gamma_{opt} \approx 4$~GHz, and optical driving amplitudes in frequency units $\Omega_L = 3.15$~MHz and 3.45~MHz for $\sigma_+$ and $\sigma_-$ polarized beam, respectively. The last values corresponds to the on-resonance Rabi frequencies \cite{Oelsner2013} introduced by the pumping beams.

We achieve a good correspondence between our model and the experimental results. It is best close to angles of zero and $\pm \pi$, where additionally to the strong light shift also a very good pumping to the dark states is achieved. Close to angles of $\pm\pi/2$, not only the observed light shift is reduced but also the population is distributed between several ground state levels. This effect reduces the magnetic resonance amplitude and, additionally, enables probing more ground state transitions both reducing the agreement between theory and experiment.

Nevertheless the model allows the reconstruction of the laser intensities $I_L$ from
\begin{equation}
\Omega_L = \frac{1}{h} \sqrt{\frac{I_L}{4cn\epsilon_0}} \braAket{J=1/2}{\left|\vec{D} \right|}{J^\prime = 1/2}.
\end{equation}
Here the vacuum permittivity $\epsilon_0$, speed of light $c$, reduced transition dipole moment $\braAket{J=1/2}{\left|\vec{D} \right|}{J^\prime = 1/2}$ \cite{Steck2010}, as well as the refractive index $n$ enter to the equation. We found the two intensities to be 64 and 76~W/m$^2$. Since their ratio does not correspond to the scaling factor $a$, we expect some influence of slightly varying beam profiles or non-perfect circular polarization. If we assume the laser power distributed equally to a circular beam with a diameter of 4~mm we find a total value of about 1~mW which fits well to separate measurements.

\subsection{The dc-photocurrent}
\begin{figure}[htb]
\includegraphics[]{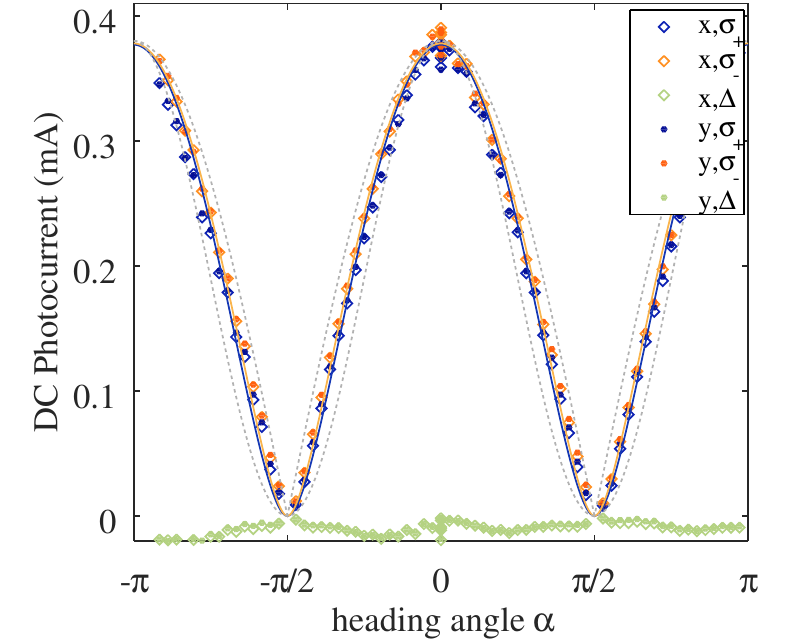}
\caption{Photocurrent as a function of the heading angle. The diamonds and dots correspond to the data recorded with the $x$ and $y$ coil producing the $B_1$ field, respectively. In the picture the difference $\Delta$ of $\sigma_-$ minus $\sigma_+$ values are added to visualize the balancing. The solid lines are calculated according to Eq.~\protect\eqref{Eq:zeropop}. The dotted gray lines correspond to the functions $0.38$~mA$\times\left| \cos \alpha \right|$ and $0.38$~mA$\times\cos^2 \alpha$ for a better visual comparison of the results to these functions.}\label{Fig:DCcurr}
\end{figure}

A second basic characteristic is found for the dc-photocurrent as introduced in Eq.~\ref{Eq:Lorentzfit}. It is plotted as a function of the heading angle in Fig.~\ref{Fig:DCcurr}.

As demonstrated in the figure, we observe a dependence that roughly follows a $\left|\cos \alpha \right|$ or $\left|\cos \alpha \right|^2$ function. Furthermore, the orientation of the $B_1$ coils has no influence to the photocurrent away from the magnetic resonance, as expected. Although we used an electronic balancing, we still observe a slightly higher dc-photocurrent for the $\sigma_-$ beams compared to the $\sigma_+$ helicity of about 2~\% in maximum that is best demonstrated from the difference signal.

The explanation of the shape of the measurement curves is found in the interaction of the dipole moment $\vec{D}$ with the laser's electric field $\vec{E}$ as given in Eq.~\eqref{Eq:dipole}. We included this effect by modified relaxation and excitation rates depending on the effective circularly polarized laser intensity \eqref{Eq:moddiss}. The measured photocurrent is increased when the pumping of the atoms to dark states is more efficient because they cannot absorb anymore light. A large dark state population is related to large differences in the ratio of light pumping to larger and smaller magnetic quantum numbers. They are respectively proportional to $D_+ = D_x + i D_y$ and $D_- = D_x - i D_y$.

Our model (Eq.~\ref{Eq:zeropop}) predicts a change from a $\cos^2\alpha$ dependence to one proportional to $\left| \cos \alpha \right|$ when the effective pumping amplitude $\Omega_L$ is increased compared to the relaxation of the polarization given by a rate $\gamma_r$. That allows for a more accurate fit of the photocurrent as a function of the orientation angle $\alpha$ as demonstrated by the solid lines in Fig.~\ref{Fig:DCcurr}. For this curve, we estimate the ratio of relaxation to optical pumping rate to be $p_1 \approx 0.3$ and the amplitudes $I_{0}=0.49$~mA. Keeping in mind the electronic amplification of the $\sigma_+$ channel, the variable $p_1$ is multiplied by $1/1.6$ for this channel making it necessary also to adjust the amplitude $I_0$ when fitting the curve measured for this helicity to 0.56~mA. Additionally, in consistency with the experiment, our model predicts a smaller photocurrent with decreasing the optical pumping amplitude $\propto\Omega_p$ that usually is not included in Bloch equations.

\subsection{Magnetic resonance amplitude}
\begin{figure}[h]
\includegraphics[]{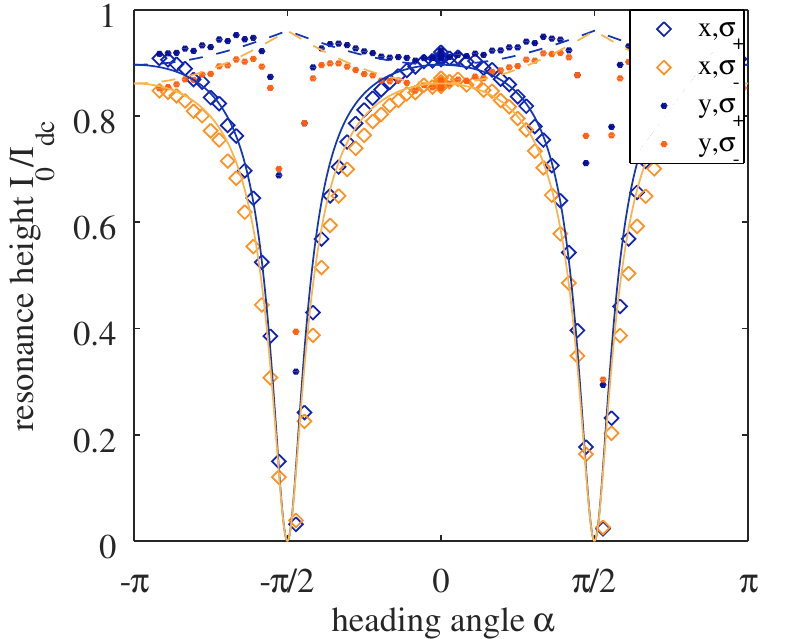}
\caption{Normalized magnetic resonance height $I_0/I_{dc}$ as a function of the heading angle. Again the diamonds and dots correspond to the $x$ and $y$ orientation of the $B_1$ field, respectively. The calculation of the solid and dashed theoretical lines for the respective $x$ and $y$ coil is explained in the main text. Their colors are adjusted to the corresponding data points.} \label{Fig:Amp}
\end{figure}

In contrast to the very similar angular dependencies of the dc-photocurrent for the two different $B_1$ coils, a clear discrepancy is found in the normalized amplitude of the magnetic resonance signal $i_0 = I_0/I_{dc}$, as demonstrated in Fig.~\ref{Fig:Amp}.

This mentioned discrepancy is well explained by the additional modification of the effective $B_1$ field amplitude when modifying its direction compared to $\vec{B}_0$. We included this in our calculation by the factor $c$ that is constantly 1, when using the $y$-coil, and $\left|\cos \alpha\right|$, when the $x$-coil is used. With Eq.~\ref{Fig:Amp} we fit the normalized amplitudes presented in Fig.~\ref{Fig:Amp} and found very good agreement between experiment and theory with a factor $p_2 = 0.12$. Note, for $\sigma_+$ we additionally had to adjust the value of $p_2$ by a factor of $1.6$. The dependence we observed for the resonance amplitude driven by the x-coil is strongly influenced by the $\cos$ function given for the effective $B_1$-field amplitude $\Omega$. Thus the amplitude values is maximal at $\alpha = 0$ and $\pi$ and tends to zero in vicinity of $\alpha = \pm\pi/2$, where effectively no $B_1$ field remains.

In contrast, when using the $y$-coil, our theory predicts an increase in resonance amplitude towards one for angles close to $\alpha = \pm \pi/2$. Except in close vicinity of these angles this general tendency is also found in our experiment. From our model, it is clear that this increase is connected to a reduction of the optical pumping to the dark state: A given strength of the $B_1$-field corresponds to a certain rate of population shifting back from the dark state to absorbing states. In other words, the $B_1$ field introduces a Rabi oscillation whose frequency depends on the field amplitude. Thus for larger powers, the shifting of population to the absorbing state is faster. If this rate is smaller than the possible population transfer that can be achieved by the pumping laser, the measured photocurrent in resonance does not reach down to zero. Therefore the amplitude $I_0 / I_{dc}$ is smaller than one. When rotating towards $\pm \pi/2$ and using the $y$-coil, the effective $B_1$ field amplitude stays constant. Because at the same time the optical pumping to the dark state gets less effective, the photocurrent amplitude is increased towards one. That means that all the atoms that are optically pumped contribute to the resonance with the $B_1$ field. Still, our model does not accurately reproduce the extracted values in close vicinity of $\alpha = \pm \pi/2$, where the experimentally observed amplitudes drop to zero. This deviation probably results from not considering the linearly polarized pumping at these angles, that not only leads to optical alignment but also takes the role of the $B_1$ field in redistributing population.

\subsection{Resonance width}

\begin{figure}[h]
\includegraphics[]{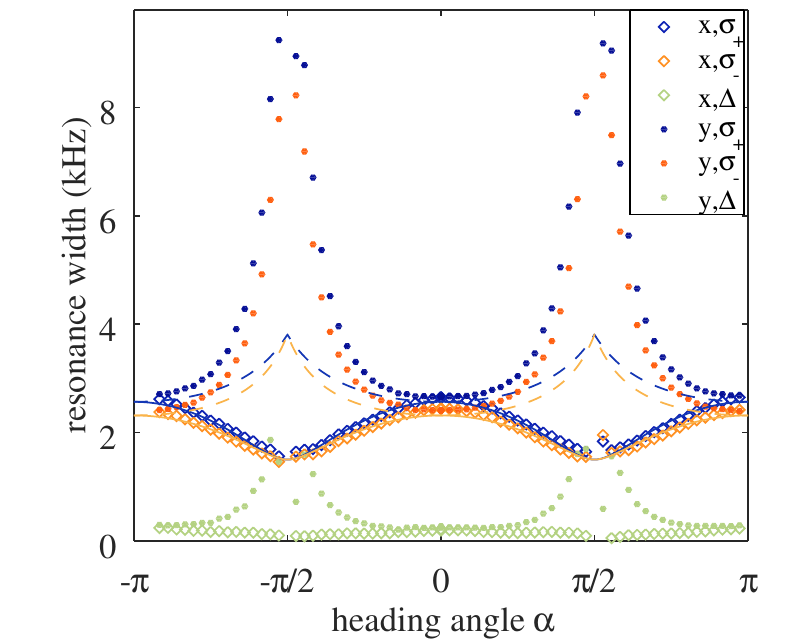}
\caption{ FWHM of the magnetic resonances as a function of the angle $\alpha$. Again diamonds and dots correspond to the use of x and y coil for applying the rf-field, respectively. In green, we added the corresponding difference between $\sigma_+$ and $\sigma_-$ result. The solid lines correspond to calculation results following Eq.~\protect\eqref{Eq:width} in the same color as the corresponding data points. } \label{Fig:width}
\end{figure}

A fitting of the experimentally observed resonance widths with the same parameters $p_1$ and $p_2$ turned out to be unsuccessful. Thus we found it necessary to introduce an additional fitting parameter $p_3$ to Eq.~\ref{Eq:width}. Theoretical curves with parameters $p_3 = 3.5$ and $\Gamma_\varphi = 350$ Hz are plotted together with the experimental data points in Fig.~\ref{Fig:width}.

Our adjusted model again fits nicely to the experimental results in the case of the $x$-coil as demonstrated by the solid lines' correspondence to the data presented as diamonds in Fig.~\ref{Fig:width}. We observe a reduction of the power broadening introduced by the rf-field. Therefore, we assume that close to angles of $\alpha = \pm \pi/2$ the minimal possible magnetic resonance width is achieved for this certain temperature and laser power.

In contrast, applying a constant effective magnetic rf-field by the $y$-coil, leads to a strong increase of the resonance width in the experiment. We account this to a large redistribution of population between Zeeman states resulting in an overlay of several ground state transitions and the linear optical pumping remaining at these angles. Since our model is restricted to two-levels and we neglected the linear pumping, we fail to catch the magnitude of this increase in magnetic resonance width. Still, we note that the qualitative behavior is accurately reproduced by the theoretical model.

As already mentioned above, we account the factor $p_3$ to a power broadening due to the strong laser power. To justify this assumption we can estimate the expected laser power broadening similar to Eq.~\ref{Eq:powerbroad1} by (compare e.g. Ref.~\cite{Loudon2000})
\begin{equation}\label{Eq:powerbroad2}
\Delta\nu_{laser} =   \frac{\Omega_L^2}{2\Gamma_\varphi \Gamma_{opt}} = 3.5,
\end{equation}
where we used the values for $\Omega_L$ of the $\sigma_+$ beam and $\Gamma_{opt}$ as noted above. Since this value is in agreement with the fitting parameter, we identify the strong power of the detuned laser as one important source of broadening of the magnetic resonance.

\subsection{Shot-noise-limited sensitivity}

\begin{figure}[h]
\includegraphics[]{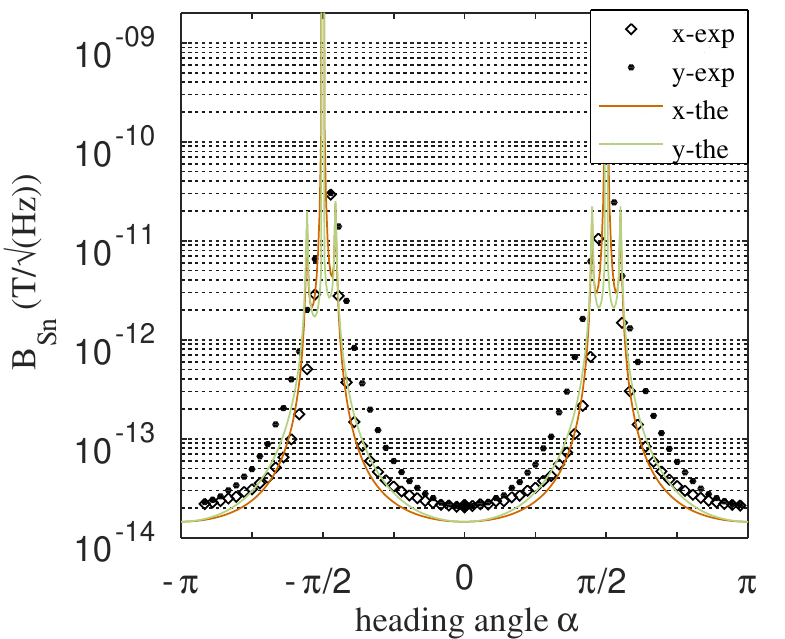}
\caption{Shot-noise-limited resolution as a function of the heading angle. The black diamonds and dots correspond to the values extracted from the experimental curves for the x and y coil, respectively. The calculated, expected dependencies in case of optimal signal balancing from the theoretical model are added as solid orange and green line.} \label{Fig:resol}
\end{figure}

Finally, as a key parameter of the proposed magnetometer we extracted the shot-noise limited sensitivity of the LSD-Mz signal as explained in Sec.~\ref{sec1}. The respective experimental values as a function of the angle $\alpha$ for the two rf-coil-orientations are presented in Fig.\ref{Fig:resol}.
Our experiment shows a minimal shot-noise limited resolution for our experimental parameters of 20~fT/$\sqrt{\text{Hz}}$. This is already close to the optimal conditions \cite{Schultze2017}. This optimal resolution is found in the case of parallel alignment of the laser direction to the magnetic field. This results from the strong dependence of the sensors performance on the pumping to the dark state and achievement of a reasonable light shift. Both of these requirements are connected to the atom-light interaction and thus are reduced towards perpendicular orientation. There, both the magnetic resonance signal as well as the light shift are reduced resulting in a loss of sensitivity.

We observe a strong dependence of the amplitude and width of the magnetic resonance signals on the angle that additionally is slightly different for the two used orientations of the $B_1$ coils. Still, because the width of the magnetic resonance as well as the light shift are both dominated by the laser intensity, the shot noise limited resolution is quite stable for angles of $\pm 20^\circ{}$ around optimal orientation.

As shown above, our model allows for a description of the orientation dependence of all the key parameters of the magnetic resonance. Thus it enables us to estimate the shot noise limited sensitivity as a function of $\alpha$ in the case of perfect balancing of the channels. To do so, we calculate the steepness of a single resonance curve as the derivative of \eqref{Eq:Lorentzfit}
\begin{equation}
\frac{s}{2} = \left|\frac{dI}{d\nu}\right| = \frac{8 I_0 \Delta\nu^2 (\nu-\nu_0)}{\left( 4\left(\nu - \nu_0 \right)^2 + \Delta \nu^2\right)^2}
\end{equation}
and evaluate it at $\nu = \nu_0 + \nu_{LS}$. There the maximal steepness of the LSD-Mz signal is achieved and corresponds to twice the value of a single magnetic resonance $s(\nu_0 + \nu_{LS}) = 2\left|\frac{dI}{d\nu}\right| (\nu_0 + \nu_{LS})$. Substituting into the equation for the shot-noise-limited resolution \eqref{Eq:Bsn} results in
\begin{equation}
B_{sn} = \sqrt{\frac{e}{I_{dc}}}\frac{1}{8\gamma i_0 }\frac{\left( 4\nu_{LS}^2 + \Delta\nu^2\right)^2}{\Delta\nu^2 \nu_{LS}},
\end{equation}
with the parameters $I_{dc}$, $i_0$, and $\Delta \nu$ defined by the respective equations \eqref{Eq:zeropop}, \eqref{Eq:ampl}, and \eqref{Eq:width}.
The theoretical estimated sensitivity is added to Fig.~\ref{Fig:resol}. In the calculation we used the experimental parameters of the starting angle and for $\sigma_+$ light and assumed perfect channel balancing. In general a slightly better sensitivity is expected from our calculation but the qualitative shape fits very well to the experiment. Note, the two additional peaks around angles of $\pm \pi/2$ result from the overlay of the two resonance curves for the $\sigma_\pm$ beams meaning $\omega_{LS} =0$, as can be seen in the plotted resonance frequencies of Fig.~\ref{Fig:freq}. This overlay results in a difference signal constantly equal to zero and thus $s\rightarrow 0$. Still, due to the lack of contrast at perpendicular configuration, this effect is not visible in the experimental data. Finally, we note that although a perfect balancing has only little influence to the sensitivity, it makes the sensor more robust against heading error, concerning the reconstructed absolute magnetic field, see green data points of Fig.~\ref{Fig:freq}.

\section*{Discussion of results and conclusion}
We experimentally analyzed the performance of an OPM operated in the LSD-Mz mode at Earth magnetic field strengths as a function of the heading of the sensor. We found that the reconstructed Larmor frequency for all heading angles corresponds accurately to the magnetic field when the two channels with different helicities are well balanced. The shot noise limited resolution strongly depends on the orientation in the external magnetic field. We demonstrated that this is related to the modified atom-light coupling responsible for both, a reduction of the photocurrent due to a reduction of spin polarization and a smaller light shift. The strong optical pumping leads to a strong contribution of power broadening to the magnetic resonance widths. This is added to the power broadening induced by the $B_1$ field. The orientation of the latter has a strong influence to the amplitude and width of the resonance signal. Nevertheless, in terms of shot noise limited resolution the two coils produce similar dependencies on the orientation angle.

Our experimental results can be explained in frame of light-shift calculations as well as Bloch equations. It was necessary to include the modified optical pumping into the latter by introducing heading dependent relaxation and excitation rates to a two-level model. This allows us to qualitatively and to a large extent also quantitatively describe our experimental results. We note, that such a model is not restricted to the description of the LSD-Mz mode, discussed in this work.

Finally, we conclude that a sensor operated in the LSD-Mz regime should be roughly $\pm 20^\circ{}$ aligned to the direction of the measured magnetic field vector to achieve a good field sensitivity. Notably, for well balanced beams the actual error in the Larmor frequency induced by light shift to each individual magnetic resonance is canceled by their subtraction. That gives a reasonable flexibility in adjusting the heading of an OPM sensor in an external magnetic field which allows their operation in Earth field strengths.
\section*{List of abbreviations}
\begin{tabular}{lcl}
BS &-& beam splitter \\
DP &-& deflecting prism\\
F &-& bandpass filter \\
FWHM &-& full width half maximum \\
HL &-& heat laser \\
KL &-& collimating lens \\
L &-& lens \\
LN &-& light narrowing \\
LP &-& linear polarizer \\
LSD-Mz &-& Light-shift dispersed M$_z$  \\
OPM &-& optically-pumped magnetometer \\
PBS &-& polarizing beam splitter \\
PD &-& photo diode \\
PL &-& pump laser \\
rf &-& radio frequency\\
\end{tabular}

\section*{Acknowledgements}
We acknowledge the support of the clean room staff under supervision of U. H\"ubner in the Center of Micro- and Nanotechnologies at Leibniz IPHT in the fabrication of the cesium vapor cell used in the experiment. We thank the Leibniz Association's Open Access Publishing Fund for financial support of this publication.

\section*{Funding}
The authors acknowledge the financial support by the Federal Ministry of Education and Research (BMBF) of Germany under Grant No. 033R130E (DESMEX). The project 2017 FE 9128, funded by the Free State of Thuringia, was co-financed by European Union funds under the European Regional Development Fund (ERDF). This work was conducted using the infrastructure supported by the Free State of Thuringia under Grant No. 2015 FGI 0008 and co-financed by European Union funds under the European Regional Development Fund (ERDF). This work has received funding from the German Research Foundation (DFG) under Grant No. SCHU 2845/2-1; AOBJ 621093.

\section*{Competing interests}
The authors declare that they have no competing interests.

\section*{Availability of data and materials}
The data generated and analysed during the current study is available from the corresponding author on reasonable request.

\section*{Author's contributions}
RIJ, VS, and RS planned the experiment. RIJ, VS, and GO created the experimental setup and carried out the measurements. GO made the data analysis and theoretical calculations. All authors participated in interpretation of results and writing the manuscript. RS supervised the project.

\bibliography{litheading}
\end{document}